\documentclass[conference]{IEEEtran}
\IEEEoverridecommandlockouts

\usepackage{cite}
\usepackage{amsmath,amssymb,amsfonts}
\usepackage{algorithmic}
\usepackage{graphicx}
\usepackage{textcomp}
\usepackage{subfigure}
\usepackage{stfloats}
\usepackage{epstopdf}
\usepackage{multirow}
\usepackage{booktabs}
\usepackage{subfigure}
\usepackage{microtype}
\usepackage{setspace}
\usepackage{color}
\usepackage{array}
\usepackage{pifont}

\def\BibTeX{{\rm B\kern-.05em{\sc i\kern-.025em b}\kern-.08em
		T\kern-.1667em\lower.7ex\hbox{E}\kern-.125emX}}

\DeclareRobustCommand*{\IEEEauthorrefmark}[1]{%
	\raisebox{0pt}[0pt][0pt]{\textsuperscript{\footnotesize\ensuremath{#1}}}}

\begin{document}
\title{A Practical Non-Stationary Channel Model for  Vehicle-to-Vehicle MIMO Communications}

\author{
\IEEEauthorblockN{Weidong~Li\IEEEauthorrefmark{1}, Qiuming Zhu\IEEEauthorrefmark{1,}\IEEEauthorrefmark{*}, Cheng-Xiang~Wang\IEEEauthorrefmark{2,}\IEEEauthorrefmark{3,}\IEEEauthorrefmark{*}, Fei~Bai\IEEEauthorrefmark{1}, Xiaomin~Chen\IEEEauthorrefmark{1}, Dazhuan~Xu\IEEEauthorrefmark{1}}

\IEEEauthorblockA{\IEEEauthorrefmark{1}The Key Laboratory of Dynamic Cognitive System of Electromagnetic Spectrum Space, \\
College of Electronic and Information Engineering, \\
Nanjing University of Aeronautics and Astronautics, Nanjing 211106, China}
\IEEEauthorblockA{\IEEEauthorrefmark{2}National Mobile Communications Research Laboratory, School of Information Science and Engineering, \\
Southeast University, Nanjing 210096, China}
\IEEEauthorblockA{\IEEEauthorrefmark{3}Purple Mountain Laboratories, Nanjing 211111, China}
\IEEEauthorblockA{\IEEEauthorrefmark{*}Corresponding author}
Email: \{liweidong, zhuqiuming\}@nuaa.edu.cn, chxwang@seu.edu.cn, cypress\_f@163.com, \\ \{chenxm402, xudazhuan\}@nuaa.edu.cn}

\maketitle

\begin{abstract}
In this paper, a practical model for non-stationary Vehicle-to-Vehicle (V2V) multiple-input multiple-output (MIMO) channels is proposed. The new model considers more accurate output phase of Doppler frequency and is simplified by the Taylor series expansions. It is also suitable for generating the V2V channel coefficient with arbitrary velocities and trajectories of the mobile transmitter (MT) and mobile receiver (MR). Meanwhile, the channel parameters of path delay and power are investigated and analyzed. The closed-form expressions of statistical properties, i.e., temporal autocorrelation function (TACF) and spatial cross-correlation function (SCCF) are also derived with the angle of arrival (AoA) and angle of departure (AoD) obeying the Von Mises (VM) distribution. In addition, the good agreements between the theoretical, simulated and measured results validate the correctness and usefulness of the proposed model.
\end{abstract}



\section{Introduction}
 V2V communication systems with MIMO technology can improve comfort in driving and reduce the economic loss in traffic by exchanging the information between different vehicles. The MIMO technology can improve the channel capacity, bandwidth efficiency and link reliability \cite{Zheng17, Liu17_SCIS}. The V2V MIMO communication system has been viewed as a significant component of intelligent transportation and an application of fifth generation (5G) communication systems \cite{WCX18_Survey, WCX18_TCom, Ge16_JSAC}. However, V2V channels are quite different from the traditional fixed-to-mobile (F2M) channels due to the moving transmitter and receiver and rapidly varying channel characteristics \cite{Makhoul17, Zhu19_WCL, Fan16, Guan16, Ge15_TCom, Zhong17_JSAC}.

 \par The geometry-based stochastic models (GBSMs) with wide-sense stationary (WSS) assumption for V2V channels have been widely accepted in the past few decades \cite{Sangjo18, Liang16_ChinaCom, Adrian15}. The statistical properties such as TACF and SCCF have been studied in \cite{Sangjo18, Liang16_ChinaCom}. However, the measured results have indicated that the WSS assumption is only suitable for short time intervals \cite{Adrian15} or short distance, i.e., 4.5m in non-line of sight (NLoS) scenarios. In other words, the V2V channels should be non-stationary and have the time-variant statistical properties under realistic communication scenarios.

 \par Recently, several V2V channel models considering the time-variant non-stationary properties have been proposed \cite{Gu18, Liang18_Access, WCX16_TVT, WCX15_TWC, Ruisi18-TWC, Jiang19_Access, WCX18_TWC, WCX_Access19, Zhu18_Access, Zhu19_CL}. Among them, the models in \cite{Gu18} and \cite{Liang18_Access} were proposed for single-input single-output (SISO) channels, where the time-variant SCCFs and TACFs were also been studied. The authors in \cite{WCX16_TVT, WCX15_TWC, Ruisi18-TWC, Jiang19_Access, WCX18_TWC} studied the V2V model and statistical properties for MIMO channels. However, the output phase of Doppler frequency in these models \cite{Gu18, Liang18_Access, WCX16_TVT, WCX15_TWC, Ruisi18-TWC, Jiang19_Access, WCX18_TWC} are not accurate compared with the theoretical ones \cite{WCX_Access19, Zhu18_Access, Zhu19_CL}. The authors in \cite{WCX_Access19} and \cite{Zhu18_Access} investigated the MIMO V2V channels but the proposed models are complicated and not practical to derive the closed-form expressions of SCCF and TACF. Note that the authors in \cite{Zhu19_CL} gave the TACF with the help of Taylor series expansions, but they only focused on SISO channels. This paper aims to fill the above research gaps.

 \par In this paper, we develop a practical non-stationary V2V MIMO channel model with arbitrary velocities and trajectories of the MT and MR. With the help of Taylor series expansions, the new model is easy to simulate and can clearly reveal the impact of velocity variations on the channel characteristics. Meanwhile, the time evolving algorithms of channel parameters, i.e., the path power and path delay, are given and analyzed. In addition, the closed-form expressions of SCCF and TACF for the proposed model are also investigated, derived, and verified by the simulated and measured results.

 \par The rest of this paper is organized as follows. In Section II, the theoretical model for V2V MIMO channels is presented. Section III gives the proposed practical model of theoretical model and the algorithms of channel
 parameters. In Section IV, the statistical properties of simulated method are investigated and derived.
 The simulation and validation are performed in Section V. Finally, the conclusions are given in Section VI.
%
%
%
%


\begin{figure*}[!b]
\normalsize
\setcounter{equation}{2}
\begin{equation}
\begin{array}{l}
{{\tilde h}_{p,q,n}}(t,\tau ) = \mathop {\lim }\limits_{M \to \infty } \frac{1}{{\sqrt M }}\sum\limits_{m = 1}^M {} {{\rm{e}}^{{\rm{j}} \cdot {\Phi _{n,m}}(t)}} \cdot {{\rm{e}}^{{\rm{j}} \cdot (2\pi  \cdot \int_0^t {} f_{n,m}^{}(t'){\rm{d}}t' + \theta _{n,m}^{})}}\\
 \ \ \ \ \ \ \ \ \ \ \ \ \ \ = \int\limits_{\alpha _n^{{\rm{MT}}} \in (0,2\pi ]} {} \int\limits_{\alpha _n^{{\rm{MR}}} \in (0,2\pi ]} {} {{\rm{e}}^{{\rm{j}} \cdot {\Phi _{n,m}}(t)}} \cdot {{\rm{e}}^{{\rm{j}} \cdot (2\pi  \cdot \int_0^t {} {f_{n,m}}(t'){\rm{d}}t' + {\theta _{n,m}})}} \cdot {p_{\alpha _{n,m}^{{\rm{MT}}}}}(\alpha _n^{{\rm{MT}}}) \cdot {p_{\alpha _{n,m}^{{\rm{MR}}}}}(\alpha _n^{{\rm{MR}}}){\rm{d}}\alpha _n^{{\rm{MT}}}{\rm{d}}\alpha _n^{{\rm{MR}}}
\end{array}
\label{3}
\end{equation}
\end{figure*}

\section{Theoretical Model for V2V MIMO Channels}
\par Let's consider a typical twin-cluster V2V communication model, where the MT and MR are moving in arbitrary trajectories with varying velocities denoted as ${{\bf{v}}^i}(t)$, $i \in \{ {\rm{MT,MR}}\} $ representing MT or MR in brief. The system includes two local coordinate systems denoted as the MT coordinate system and MR coordinate system, and their origins are at the central position of MT and MR, respectively. In V2V communication scenarios, the location of the MT or MR antenna element can be denoted as ${{\mathbf{d}}^{i}}={{\left[ d_{x}^{i},d_{y}^{i} \right]}^{\text{T}}}$, $i \in \{ {\rm{MT,MR}}\} $, where $d_x^i$ and $d_y^i$ represent the locations on the $x$ axis and $y$ axis. The V2V channel includes the multiple-bounced propagation path, which contains the Line-of-Sight (LoS) when there is no clusters existing in the certain path. Under NLoS scenario, the clusters in twin-cluster model can be divided into three parts \cite{Hofstetter06_EuCAP}, i.e., the first one ${\bf{S}}_n^{{\rm{MT}}}$, the last one ${\bf{S}}_n^{{\rm{MR}}}$, and the rest between ${\bf{S}}_n^{{\rm{MT}}}$ and ${\bf{S}}_n^{{\rm{MR}}}$. Only the first and last clusters modeled by their velocities and locations, while the rest can be abstracted by a virtual link \cite{Zhu18_TWC}.
\begin{figure}[bt]
 {\includegraphics[width=80mm,height=58mm]{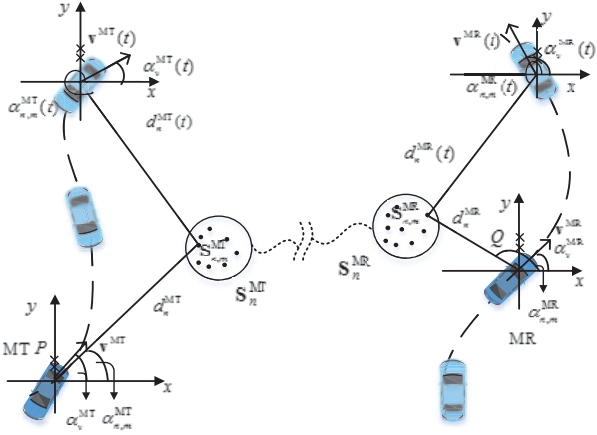}}
\caption{Typical twin-cluster V2V communication scenario.}
\label{fig:1}
\end{figure}
\par Under the general scattering environment, the V2V multiple-input multiple-output (MIMO) channel between the MT and MR can be expressed as a $P \times Q$ complex matrix, i.e.
\setcounter{equation}{0}
\begin{equation}
{\bf{H}}(t,\tau ) = {\left[ {{h_{p,q}}(t,\tau )} \right]_{P \times Q}}
\label{1}
\end{equation}
\noindent where ${{h}_{p,q}}(t,\tau )$ denotes the complex channel impulse response (CIR) between the MT antenna element $p$ $(p=1,2,\ldots ,P)$ and MR antenna element $q\text{ }(q=1,2,\ldots ,Q)$, and it can be further expressed as \cite{Zhu18_Access}
\setcounter{equation}{1}
\begin{equation}
{{h}_{p,q}}(t,\tau )=\sum\limits_{n=1}^{N}{{{P}_{n}}(t){{{\tilde{h}}}_{p,q,n}}(t)}\delta (\tau -{{\tau }_{n}}(t))
\label{2}
\end{equation}
\noindent where $N$ denotes the number of propagation paths, ${{P}_{n}}(t)$ denotes the path power, ${{\tau }_{n}}(t)$ represents the time delay, ${{\tilde{h}}_{p,q,n}}(t)$ represents the complex channel gain as (\ref{3}), where ${{\Phi }_{n,m}}(t)={{\mathbf{\tilde{s}}}^{\text{MT}}}(t)\cdot \mathbf{d}_{p}^{\text{MT}}+{{\mathbf{\tilde{s}}}^{\text{MR}}}(t)\cdot \mathbf{d}_{q}^{\text{MR}}$ denotes the phase related with the antennas index and scattering environment, where ${{\mathbf{\tilde{s}}}^{i}}(t)=\left[ \cos (\alpha _{n,m}^{i}(t)),\sin(\alpha _{n,m}^{i}(t)) \right]$, $i \in \{ {\rm{MT,MR}}\} $ denotes the arrival or departure unit direction vectors, $\mathbf{d}_{p}^{\text{MT}}={{\left[ d_{p,x}^{\text{MT}},d_{p,y}^{\text{MT}} \right]}^{\text{T}}}$ and $\mathbf{d}_{q}^{\text{MR}}={{\left[ d_{q,x}^{\text{MR}},d_{q,y}^{\text{MR}} \right]}^{\text{T}}}$ denote the position vector of the $p$th MT antenna or the $q$th MR antenna. In (\ref{3}), $\alpha _{n,m}^{i}(t)$ represents the angle of arrival (AoA) or angle of departure (AoD), ${{p}_{\alpha _{n,m}^{i}}}(\alpha _{n}^{i})$ denotes the probability density function (PDF) of AoA or AoD, and the AoA and AoD are usually independent when the clusters are independent \cite{Zhu19_WCL}. In (3), ${{f}_{n,m}}(t)$ represents the Doppler frequency.

\begin{figure*}[!b]
\normalsize
\setcounter{equation}{12}
\begin{equation}
d_{n}^{i}(t)=\sqrt{{{\left( d_{n}^{i}\cos(\bar{\alpha }_{n}^{i})-\int_{0}^{t}{(v_{0}^{i}+a_{0}^{i}t')\cdot \cos (\alpha _{v}^{i}+b_{0}^{i}t')\text{d}t'} \right)}^{2}}+{{\left( d_{n}^{i}\sin (\bar{\alpha }_{n}^{i})-\int_{0}^{t}{(v_{0}^{i}+a_{0}^{i}t')\cdot \sin (\alpha _{v}^{i}+b_{0}^{i}t')\text{d}t'} \right)}^{2}}}
\label{13}
\end{equation}

\setcounter{equation}{14}
\begin{equation}
{{\tau }_{n}}(t)\approx \frac{d_{n}^{\text{MT}}+d_{n}^{\text{MR}}-(v_{0}^{\text{MT}}\cdot \cos (\bar{\alpha }_{n}^{\text{MT}}-\alpha _{v}^{\text{MT}})+v_{0}^{\text{MR}}\cdot \cos (\bar{\alpha }_{n}^{\text{MR}}-\alpha _{v}^{\text{MR}}))\cdot t}{\text{c}}+{{\tilde{\tau }}_{n}}(t)
\label{15}
\end{equation}

\setcounter{equation}{21}
\begin{equation}
\begin{array}{l}
   \rho _{n}^{i}(t,\Delta t)=\int_{-\pi }^{\pi }{{}}{{\text{e}}^{-\text{j}\cdot (\frac{2\pi }{3\lambda }\cdot a_{0}^{i}\cdot k_{0}^{i}\cdot ({{(t+\Delta t)}^{3}}-{{t}^{3}})+\frac{\pi }{\lambda }\cdot (a_{0}^{i}\cdot \cos(\alpha _{n,m}^{i}-\alpha _{v}^{i})+v_{0}^{i}\cdot k_{0}^{i})\cdot ({{(t+\Delta t)}^{2}}-{{t}^{2}})+\frac{2\pi }{\lambda }\cdot v_{0}^{i}\cdot \cos(\alpha _{n,m}^{i}-\alpha _{v}^{i})\cdot \Delta t)}} \\
  \ \ \ \ \ \ \ \ \ \ \ \ \ \cdot \sqrt{{{p}_{\alpha _{n,m}^{i}}}(\alpha _{n}^{i}(t))}\sqrt{{{p}_{\alpha _{n,m}^{i}}}(\alpha _{n}^{i}(t+\Delta t))}\text{d}\alpha _{n}^{i} \\
\end{array}
\label{22}
\end{equation}

\setcounter{equation}{22}
\begin{equation}
\rho _{n}^{i}(t,\Delta t)\ =\frac{{{\text{I}}_{0}}(\sqrt{({{\kappa }^{i}}^{2}-R{{_{n}^{i}}^{2}}-S{{_{n}^{i}}^{2}}+2\text{j}\cdot {{\kappa }^{i}}\cdot R_{n}^{i}\cdot \text{cos(}\alpha _{v}^{i}(t)-\bar{\alpha }_{n}^{i}(t)\text{)}-2\text{j}\cdot {{\kappa }^{i}}\cdot S_{n}^{i}\cdot \text{sin(}\alpha _{v}^{i}(t)-\bar{\alpha }_{n}^{i}(t)\text{)})})}{{{\operatorname{I}}_{0}}({{\kappa }^{i}})}
\label{23}
\end{equation}
\end{figure*}

\section{Computation Methods for Time-variant Parameters}
\subsection{Proposed channel model}
\par It should be mentioned that the theoretical model is suitable for the V2V channels with time-variant velocities, but it's not practical to simulate the model and calculate the time-variant channel parameters. The initial parameters and moving trajectories of two terminals can be provided by users in practice, while theoretical model is more general and cannot be applied with practical parameters directly. In order to simulate the V2V channels, we firstly rewrite the complex channel gain ${{\tilde{h}}_{p,q,n}}(t,\tau )$ as
\setcounter{equation}{3}
\begin{equation}
{{\tilde{h}}_{p,q,n}}(t,\tau )=\!\!\frac{1}{\sqrt{M}}\sum\limits_{m=1}^{M}{{}}{{\text{e}}^{\text{j}\cdot (2\pi \cdot \int_{0}^{t}{{}}{{f}_{n,m}}(t')\text{d}t'+{{\Phi }_{n,m}}(t)+{{\theta }_{n,m}})}}
\!\!\!\!\!\!\label{4}
\end{equation}
\noindent where ${{\theta}_{n,m}}$ represents the initial random phase and distributed uniformly over $(0,2\pi]$. Since the terminals are usually independent from each other, the total Doppler frequency can be calculated by ${{f}_{n,m}}(t)=f_{n,m}^{\text{MT}}(t)+f_{n,m}^{\text{MR}}(t)$, where $f_{n,m}^{i}(t)$, $i \in \{ {\rm{MT,MR}}\} $ denotes the Doppler frequency of the MT or MR and can be expressed as
\setcounter{equation}{4}
\begin{equation}
f_{n,m}^i(t) = \;\frac{{{v^i}(t)}}{\lambda } \cdot \cos (\alpha _{n,m}^i(t) - \alpha _v^i(t))
\!\!\!\!\!\!\label{5}
\end{equation}
\noindent where ${v^i}(t)$ and $\alpha _v^i(t)$ represent the speed and movement direction of MT or MR, respectively. Under V2V communication scenarios, the time-variant speed and moving direction can be assumed to change linearly, i.e., ${v^i}(t) = v_0^i + a_0^i \cdot t$ and $\alpha _v^i(t) = \alpha _v^i + {b_0^i} \cdot t$, where $v_0^i$ and $\alpha _v^i$ mean the initial values of speed and moving direction of MT or MR, $a_0^i$ and $b_0^i$ represent the accelerations of speed and moving direction of MT or MR, respectively. Then, we expand $\cos (\alpha _{n,m}^i(t) - \alpha _v^i(t))$  by Taylor series expansion at $t=0$ and take the first two terms to approximate it as
\setcounter{equation}{5}
\begin{equation}
\cos (\alpha _{n,m}^i(t) - \alpha _v^i(t)) = \cos (\alpha _{n,m}^i - \alpha _v^i) + k_0^i \cdot t
\!\!\!\!\!\!\label{6}.
\end{equation}
\noindent where $\alpha _{n,m}^i$, $i \in \{ {\rm{MT,MR}}\} $ denotes the initial AoA or AoD, $k_0^i$ represents the coefficient of first order in Taylor series expansion and can be expressed as
\setcounter{equation}{6}
\begin{equation}
k_{0}^{i}=-\frac{v_{0}^{i}\cdot {\sin^{2}}(\alpha _{n,m}^{i}-\alpha _{v}^{i})}{d_{n}^{i}}+b_{0}^{i}\cdot \sin(\alpha _{n,m}^{i}-\alpha _{v}^{i})
\!\!\label{7}
\end{equation}
\noindent where $d_n^i$, $i \in \{ {\rm{MT,MR}}\} $ denotes the initial distance between the terminal MT or MR and cluster ${\bf{S}}_n^{{\rm{MT}}}$ or ${\bf{S}}_n^{{\rm{MR}}}$, and $\alpha _{n,m}^i$, $i \in \{ {\rm{MT,MR}}\} $ represents the initial AoA or AoD at $t = 0$. It should be mentioned that only the two terms of Taylor series expansion can match the $\cos (\alpha _{n,m}^i(t) - \alpha _v^i(t))$ well. Finally, $f_{n,m}^{i}(t)$ can be simplified and obtained by
\setcounter{equation}{7}
\begin{equation}
\begin{array}{l}
   \!\!\!\!f_{n,m}^{i}(t)\!\approx\! \frac{a_{0}^{i}{{k}_{0}}^{i}}{\lambda }{{t}^{2}}\!\!+\!\!\frac{a_{0}^{i}\cdot \cos (\alpha _{n,m}^{i}\!-\!\alpha _{v}^{i})+v_{0}^{i}k_{0}^{i}}{\lambda }t \!+\!\frac{v_{0}^{i}\cos (\alpha _{n,m}^{i}\!-\!\alpha _{v}^{i})}{\lambda } \\
\end{array}.
\!\!\!\!\!\!\label{8}
\end{equation}
\par Similarly, the phase ${{\Phi }_{n,m}}(t)$ in (\ref{4}) can be calculated by ${{\Phi }_{n,m}}(t)=\Phi _{n,m}^{\text{MT}}(t)+\Phi _{n,m}^{\text{MR}}(t)$, where $\Phi _{n,m}^{i}(t)$, $i \in \{ {\rm{MT,MR}}\} $ denotes the phase at the MT or MR. By using Taylor series expansions, the phase $\Phi _{n,m}^i(t)$ can be simplified and obtained by
\setcounter{equation}{8}
\begin{equation}
\begin{array}{l}
   \Phi _{n,m}^{i}(t)\approx \frac{2\pi }{\lambda }(k_{1}^{i}d_{u,x}^{i}+k_{2}^{i}d_{u,y}^{i})t \\
  \ \ \ \ \ \ \ \ \ \ \ +\frac{2\pi }{\lambda }(\cos (\alpha _{n,m}^{i})\cdot d_{u,x}^{i}+\sin (\alpha _{n,m}^{i})\cdot d_{u,y}^{i}) \\
\end{array}
\!\!\label{9}
\end{equation}
\noindent where  $u\in \{p,q\}$ denotes the $p$th MT antenna element or $q$th MR antenna element, $k_{1}^{i}$ and $k_{2}^{i}$ can be expressed as
\setcounter{equation}{9}
\begin{equation}
\begin{array}{l}
   k_{1}^{i}=-\frac{v_{0}^{i}}{d_{n}^{i}}\cdot {{\sin }^{2}}(\alpha _{n,m}^{i})+b_{0}^{i}\cdot \sin (\alpha _{n,m}^{i}) \\
  k_{2}^{i}=-\frac{v_{0}^{i}}{d_{n}^{i}}\cdot {{\cos }^{2}}(\alpha _{n,m}^{i})+b_{0}^{i}\cdot \cos (\alpha _{n,m}^{i}). \\
\end{array}
\label{10}
\end{equation}
\par By substituting (\ref{8}) and (\ref{9}) into (\ref{4}), ${{\tilde{h}}_{p,q,n}}(t,\tau )$ of our simulation model can be simplified as
\setcounter{equation}{10}
\begin{equation}
h_{p,q,n}^{{}}(t,\tau )=\frac{1}{\sqrt{M}}\sum\limits_{m=1}^{M}{{}}{{\text{e}}^{\text{j}\cdot (A\cdot {{t}^{3}}+B\cdot {{t}^{2}}+C\cdot t+D+\theta _{n,m}^{{}})}}
\label{11}
\end{equation}
\noindent where
\setcounter{equation}{11}
\begin{subequations}
\begin{equation}
\setlength{\abovedisplayskip}{10pt}
\setlength{\belowdisplayskip}{5pt}
A=\frac{2\pi }{3\lambda }a_{0}^{i}k_{0}^{i}
\!\!\!\!\!\!\!\!\!\label{12a}
\end{equation}
\begin{equation}
\setlength{\abovedisplayskip}{5pt}
\setlength{\belowdisplayskip}{5pt}
B=\frac{\pi }{\lambda }(a_{0}^{i}\cdot \cos(\alpha _{n,m}^{i}-\alpha _{v}^{i})+v_{0}^{i}k_{0}^{i})
\!\!\!\!\label{12b}
\end{equation}
\begin{equation}
\setlength{\abovedisplayskip}{5pt}
\setlength{\belowdisplayskip}{5pt}
C=\frac{2\pi }{\lambda }(v_{0}^{i}\cdot \cos(\alpha _{n,m}^{i}-\alpha _{v}^{i})+k_{1}^{i}d_{u,x}^{i}+k_{2}^{i}d_{u,y}^{i})
\!\!\!\!\label{12c}
\end{equation}
\begin{equation}
\setlength{\abovedisplayskip}{5pt}
\setlength{\belowdisplayskip}{5pt}
D=\frac{2\pi }{\lambda }(\cos (\alpha _{n,m}^{i})\cdot d_{u,x}^{i}+\sin (\alpha _{n,m}^{i})\cdot d_{u,y}^{i}).
\!\!\!\!\label{12d}
\end{equation}
\end{subequations}

\subsection{Time-variant parameters}
\par The time-variant distance $d_{n}^{i}(t)$ between the terminals and clusters can be calculated according to the initial values and be expressed as (\ref{13}), where $\bar{\alpha }_{n}^{i}$, $i \in \{ {\rm{MT,MR}}\} $ means the initial mean angles of AoA or AoD. By expanding $d_{n}^{i}(t)$ with the Taylor expansion at $t=0$ and only taking the first two terms, $d_{n}^{i}(t)$ can be simplified as
\setcounter{equation}{13}
\begin{equation}
d_{n}^{i}(t)\approx d_{n}^{i}-v_{0}^{i}\cdot \cos (\bar{\alpha }_{n}^{i}-\alpha _{v}^{i})\cdot t.
\label{14}
\end{equation}
\noindent The delay of the $n$th path can be divided into three parts, i.e., the delay of the MT, the delay of virtual link, and the delay of the MR. Then, it can be calculated as (\ref{15}), where $\text{c}$ denotes the speed of light, and ${{\tilde{\tau }}_{n}}(t)$ is the equivalent delay of virtual link and can be updated by a first-order filtering method \cite{Zhu18_TWC}.

\par The power of the $n$th path can be calculated as \cite{WCX18_TWC}
\setcounter{equation}{15}
\begin{equation}
{{P}_{n}}'(t)={{\text{e}}^{-{{\tau }_{n}}(t)\cdot \frac{{{r}_{\tau }}-1}{{{r}_{\tau }}\cdot {{\sigma }_{\tau }}}}}\cdot {{10}^{-\frac{{{Z}_{n}}}{10}}}
\label{16}
\end{equation}
\noindent where ${{r}_{\tau }}$, ${{\sigma }_{\tau }}$ and ${{Z}_{n}}$ denote the shadow term, delay distribution and delay spread, respectively. The normalized process can be expressed as
\setcounter{equation}{16}
\begin{equation}
{{P}_{n}}(t)=\frac{{{P}_{n}}'(t)}{\sum\nolimits_{n=1}^{N}{{}}{{P}_{n}}'(t)}.
\label{17}
\end{equation}

\section{Statistical Properties of proposed model}
\par The normalized spatial-temporal correlation function (STCF) between the ${{\tilde{h}}_{{{p}_{1}},{{q}_{1}},n}}(t)$ and ${{\tilde{h}}_{{{p}_{2}},{{q}_{2}},n}}(t+\Delta t)$ can be defined as \cite{Zhu18_Access}
\setcounter{equation}{17}
\begin{equation}
\begin{array}{l}
   \!\!\!\!\!{{\rho }_{n}}(t,\!\Delta t,\!\Delta {{d}_{p}},\!\Delta {{d}_{q}})\! =\!\!\frac{\text{E}\left[ {{{\tilde{h}}}_{{{p}_{1}},{{q}_{1}},n}}(t)\cdot \tilde{h}_{{{p}_{2}},{{q}_{2}},n}^{*}(t+\Delta t) \right]}{\sqrt{\text{E}\left[ {{\left| {{{\tilde{h}}}_{{{p}_{1}},{{q}_{1}},n}}(t) \right|}^{2}} \right]\cdot \text{E}\left[ \left| {{{\tilde{h}}}_{{{p}_{2}},{{q}_{2}},n}}(t+\Delta t) \right| \right]}} \\
\end{array}
\!\!\!\!\label{18}
\end{equation}
\noindent where $\text{E}\left[ \cdot  \right]$ denotes the expectation function, ${{(\cdot )}^{*}}$ is complex conjugate, $\Delta t$ means the time lag, $\Delta {{d}_{p}}=\left\| \mathbf{d}_{{{p}_{1}}}^{\text{MT}}-\mathbf{d}_{{{p}_{2}}}^{\text{MT}} \right\|$ and $\Delta {{d}_{q}}=\left\| \mathbf{d}_{{{q}_{1}}}^{\text{MR}}-\mathbf{d}_{{{q}_{2}}}^{\text{MR}} \right\|$ mean antenna element spacing of MT and MR, respectively. In addition, the random parameters $\alpha _{n,m}^{\text{MT}}$ and $\alpha _{n,m}^{\text{MR}}$ can be characterized by a certain PDF. Several previous work have mentioned that the AoA and AoD may follow the uniform, Gaussian, and so on under different scattering scenarios. However, the measurement results in \cite{Pedersen00} and \cite{Zajic08} revealed that the von Mise (VM) distribution is flexible and can approximate these distributions well. Thus, it is assumed in this paper that the AoA and AoD obey the VM distribution as
\setcounter{equation}{18}
\begin{equation}
{{p}_{\alpha _{n,m}^{i}}}(\alpha _{n}^{i}(t))=\frac{\exp ({{\kappa }^{i}}\cdot \cos(\alpha _{n,m}^{i}(t)-\bar{\alpha }_{n}^{i}(t)))}{2\pi \cdot {{\text{I}}_{0}}({{\kappa }^{i}})}
\label{19}
\end{equation}
\noindent where ${{\kappa }^{i}}$ denotes the factor related to the concentration of distribution, ${{\text{I}}_{0}}(\cdot )$ represents the zeroth-order modified Bessel function of the first kind, and $\bar{\alpha }_{n}^{i}(t)$ denotes the mean value of AoA or AoD.

\begin{figure*}[hb]
\setcounter{equation}{23}
\begin{subequations}
\begin{equation}
\setlength{\abovedisplayskip}{10pt}
\setlength{\belowdisplayskip}{5pt}
R_{n}^{i}=\frac{2\pi }{\lambda }\left( -\frac{v_{{}}^{i}(t)\Delta t}{2}-\frac{a_{0}^{i}\Delta {{t}^{2}}}{6}-\left( \frac{v_{{}}^{i}(t) \Delta t}{2}+\frac{a_{0}^{i} \Delta {{t}^{2}}}{3} \right)\cdot \cos \left( \left( \frac{v_{0}^{i}\cdot \sin (\bar{\alpha }_{n}^{i}-\alpha _{v}^{i})}{d_{n}^{i}}-b_{0}^{i} \right) \Delta t \right) \right)
\!\!\!\!\!\!\!\!\!\label{24a}
\end{equation}
\begin{equation}
\setlength{\abovedisplayskip}{5pt}
\setlength{\belowdisplayskip}{5pt}
S_{n}^{i}=\frac{2\pi }{\lambda }\left( \frac{a_{0}^{i}\Delta {{t}^{2}}}{3}+\frac{v_{{}}^{i}(t)\Delta t}{2} \right)\cdot \sin \left( \left( \frac{v_{0}^{i}\cdot \sin (\bar{\alpha }_{n}^{i}-\alpha _{v}^{i})}{d_{n}^{i}}-b_{0}^{i} \right)\Delta t \right)
\!\!\!\!\label{24b}
\end{equation}
\end{subequations}
\end{figure*}

\subsection{Time-variant SCCF}
\par The SCCF can be viewed as a special case of STCF with the time lag $\Delta t$ equal zero. Considering the condition that the clusters around the MT and MR are independent, the SCCF of our simulation model can be rewritten as ${{\rho }_{n}}(t,\Delta {{d}_{p}},\Delta {{d}_{q}})=\rho _{n}^{\text{MT}}(t,\Delta {{d}_{p}})\cdot \rho _{n}^{\text{MR}}(t,\Delta {{d}_{q}})$, where $\rho _{n}^{i}(t,\Delta {{d}_{u}})$, $i \in \{ {\rm{MT,MR}}\} $, $u\in \{p,q\}$ represents the SCCF at the MT or MR. In order to simplify the analyze, we consider the uniform linear array and set the antenna elements along the $y$ axis. Then, the antenna element of the MT or MR on the $x$ axis equal to zero, i.e., $d_{{{u}_{1}},x}^{i}=d_{{{u}_{2}},x}^{i}=0$, $i \in \{ {\rm{MT,MR}}\} $, $u\in \{p,q\}$. Holding this condition, the $\rho _{n}^{i}(t,\Delta {{d}_{u}})$ can be expressed as
\setcounter{equation}{19}
\begin{equation}
\rho _{n}^{i}(t,\Delta {{d}_{u}})\!=\!\!\int_{-\pi }^{\pi }{{}}\!\!{{\text{e}}^{\text{j}\cdot \frac{2\pi }{\lambda }\cdot \sin (\alpha _{n,m}^{i}(t))\cdot \Delta {{d}_{u}})}}\!\cdot \!{{p}_{\alpha _{n,m}^{i}}}(\alpha _{n}^{i}(t))\text{d}\alpha _{n}^{i}.
\!\!\!\label{20}
\end{equation}
\noindent Combining (\ref{19})--(\ref{20}) and using the integral formula in [29. eq. (3.338-4)], the closed-form of SCCF can be proved as
\setcounter{equation}{20}
\begin{equation}
\begin{array}{l}
\!\!\!\!\rho _{n}^{i}(t,\Delta {{d}_{u}})\!=\!\frac{{{\text{I}}_{0}}(\sqrt{{{\kappa }^{i}}^{2}-{{(\frac{2\pi }{\lambda }\cdot \Delta {{d}_{u}})}^{2}}+\frac{4\pi }{\lambda }\cdot \text{j}\cdot \Delta {{d}_{u}}\cdot {{\kappa }^{i}}\cdot \cos (\bar{\alpha }_{n}^{i}(t))})}{{{\text{I}}_{0}}({{\kappa }^{i}})}.
\end{array}
\!\!\!\!\label{21}
\end{equation}
\noindent Finally, the closed-form theoretical expression of SCCF for our proposed method can be obtained.

\subsection{Time-variant TACF}
\par The TACF can be reduced from STCF by setting the space of antenna elements equal zero and it can be rewritten as ${{\rho }_{n}}(t,\Delta t)=\rho _{n}^{\text{MT}}(t,\Delta t)\cdot \rho _{n}^{\text{MR}}(t,\Delta t)$, where $\rho _{n}^{i}(t,\Delta t)$, $i \in \{ {\rm{MT,MR}}\} $ denotes the TACF at the MT or MR as (\ref{22}). In (\ref{19}), when ${{\kappa }^{i}}$ is fixed or almost constant during a short time interval, the shape of distribution maintain constant and the relative angle offset is almost constant at small instants, i.e., $\alpha _{n}^{i}(t+\Delta t)-\bar{\alpha }_{n}^{i}(t+\Delta t)\approx \alpha _{n}^{i}(t)-\bar{\alpha }_{n}^{i}(t)$. Holding this condition and substituting (\ref{19}) into (\ref{22}) and using [29. eq. (3.338-4)], the closed-form of TACF can be obtained as (\ref{23}), where $R_n^i$ and $S_n^i$ can be expressed as (24).

\section{Results and Discussions}
\par In this section, three typical V2V scenarios with trajectories variations of the MT and MR are adopted to validate the proposed model by comparing the theoretical and measured results. The three scenarios are the opposite direction I, opposite direction II and right turn, which have the same speed and initial directions of the MT or MR. The first and second scenarios are different in the acceleration of speed $a_{0}^{i}$, and $a_{0}^{\text{MT}}=0$ and $a_{0}^{\text{MR}}$ set to be a constant value. The first and third scenarios are different in the acceleration of moving direction $b_{0}^{i}$, and $b_{0}^{\text{MT}}=0$ and $b_{0}^{\text{MR}}$ set to be a constant value. The VM distribution is used to approximate the PDFs of AoA and AoD and the factor of VM distribution $\kappa $ equals to 1. The carrier frequency is 2.48 GHz. The detailed velocity parameters of the MT and MR are listed in Table I.
\par By using (\ref{18})-(\ref{21}), the theoretical and simulated SCCFs of our simulated method at three instants, i.e., $t=0\text{s, }2\text{s}$ and $5\text{s}$ under three scenarios are shown in Fig. 2(a), (b) and (c), respectively. It is shown that the simulated results of our method match well with the theoretical ones, which verifies the correctness of our simulated method and our derivations. It is easy to obtain that the larger antenna spacing,
the less SCCFs. By comparing the SCCFs in Fig. 2(a) and (b), the SCCFs under opposite direction II change faster than the ones under the first scenario for the different initial value of $a_{0}^{i}$.
By comparing the SCCFs in Fig. 2(a) and (c), the SCCFs under right turn change faster than the ones under opposite direction for the different initial value of $b_{0}^{i}$.
\begin{figure}[bt]
\subfigure[] {\includegraphics[width=80mm,height=55mm]{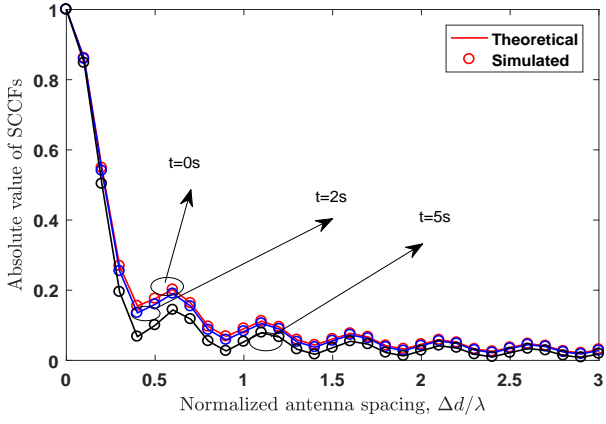}}
\hspace{0.5cm}
\subfigure[] {\includegraphics[width=80mm,height=55mm]{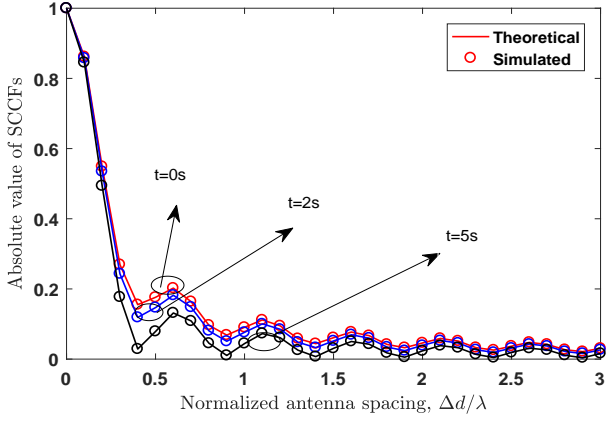}}
\hspace{0.5cm}
\subfigure[] {\includegraphics[width=80mm,height=55mm]{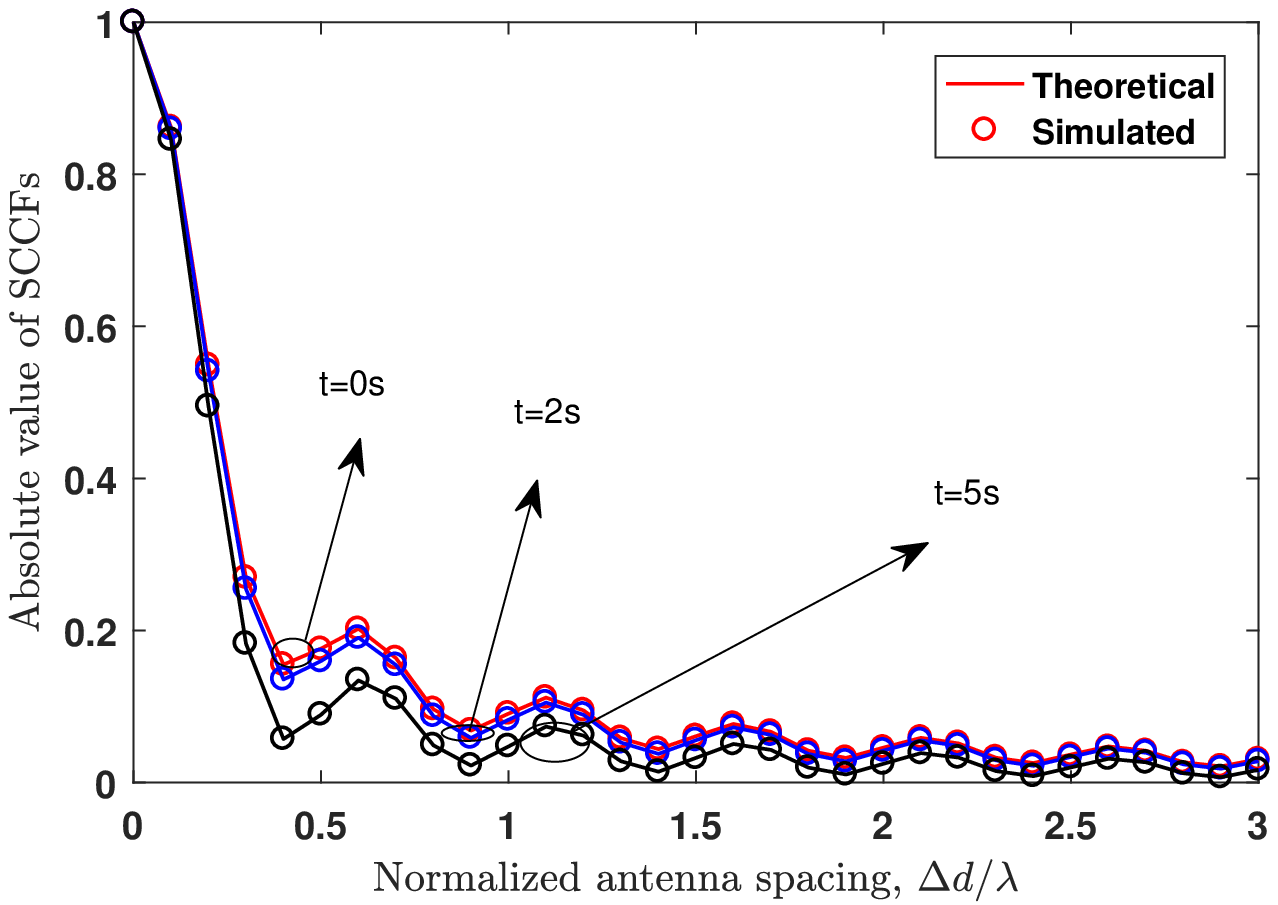}}
\caption{Theoretical and simulated SCCFs under (a) opposite direction I, (b) opposite direction II and (c) right turn at different time instants.}
\label{fig:2}
\end{figure}
\par By using (\ref{18}) and (\ref{22})-(24), the theoretical, simulated, and measured TACFs at three time instant
i.e., $t=0\text{s, 2s}$ and $\text{5s}$ are shown in Fig. 3. By comparing the TACFs under three different scenarios, the different TACFs at the same time instant indicate that the impact of parameter $a_{0}^{i}$ and $b_{0}^{i}$ having on TACFs' fading
speed. It is reasonable to be obtained that the larger the parameter, the more complicated the communication scenarios,
and the faster the speed of TACFs fading. In addition, the good agreement between the measurement \cite{Makhoul17} and
simulated results of our proposed simulated method validates the usefulness of our simulated method in actual V2V communication scenarios.
In addition, it respectively takes 4.57s and 3.82s to obtain the results of TACF and SCCF in the theoretical model with
computer configured with i5-4210M CPU and 4 GB RAM, while it takes only 1.53s and 0.86s in our simulation model with the same configuration.
\begin{figure}[bt]
\subfigure[] {\includegraphics[width=80mm,height=58mm]{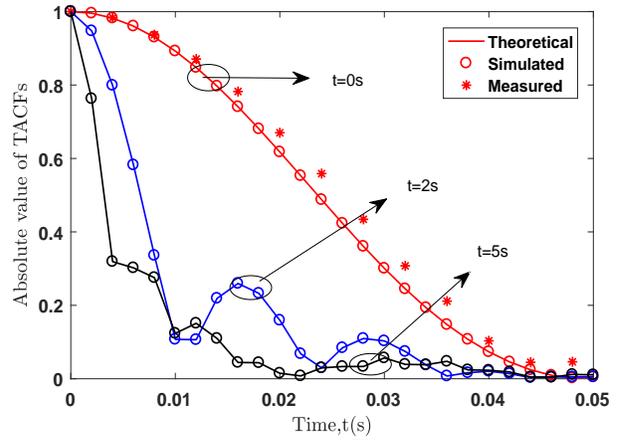}}
\hspace{0.5cm}
\subfigure[] {\includegraphics[width=80mm,height=58mm]{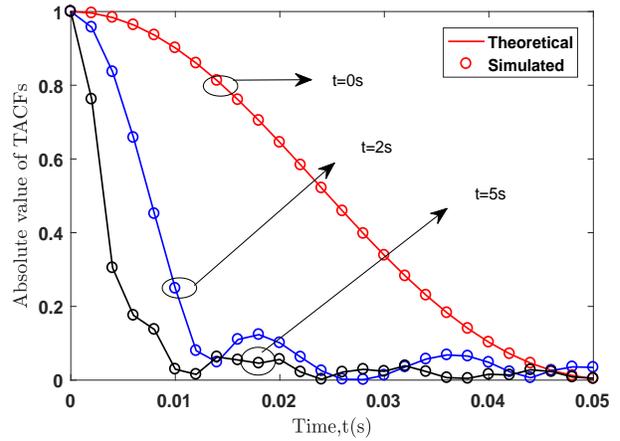}}
\hspace{0.5cm}
\subfigure[] {\includegraphics[width=80mm,height=58mm]{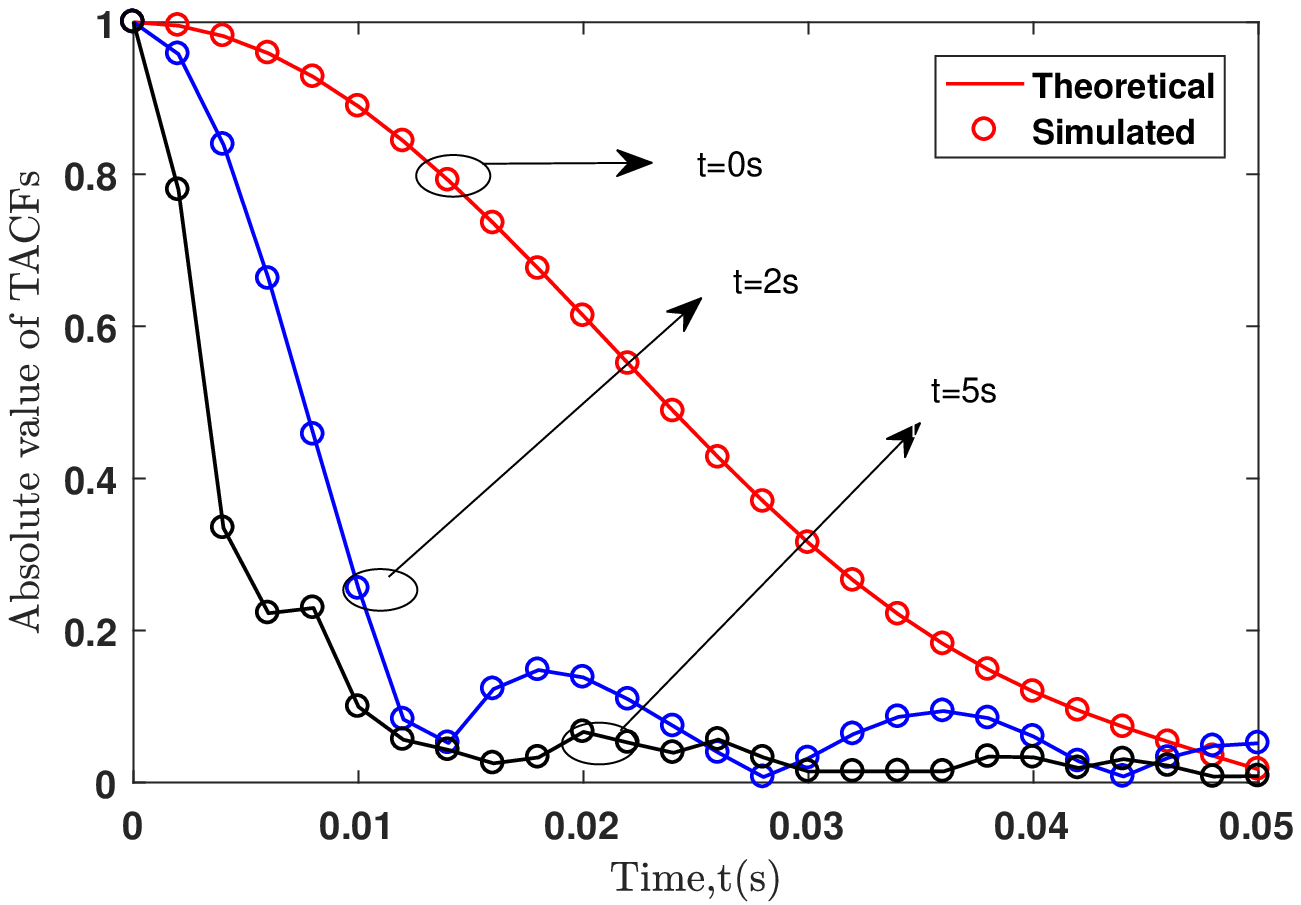}}
\caption{Theoretical, simulated, and measured TACFs under (a) opposite direction I, (b) opposite direction II and (c) right turn at different time instants.}
\label{fig:3}
\end{figure}

\section{Conclusions}
\par In this paper, we have proposed a practical model for the V2V channels considering the velocity variations and arbitrary trajectories. By using the Taylor series expansions, the channel parameters can be simplified by the initial values of velocity and location, i.e., distance, antenna spacing, etc. With the method of Taylor series expansions, the closed-form expressions of statistical properties such as the SCCF and TACF have also been derived. The correctness and usefulness of proposed model have been validated by the good agreements between the theoretical, simulated, and measured results. In the future, the proposed model can be applied on designing, evaluating, and validating the V2V communication systems under realistic scenarios.
\begin{center}
ACKNOWLEDGMENTS
\end{center}
\par This work was supported by the National Key R\&D Program of China (No.~2018YF1801101), the National Key Scientific Instrument and Equipment Development Project (No.~61827801), the National Natural Science Foundation of China (No.~61960206006), the Research Fund of National Mobile Communications Research Laboratory, Southeast University (No.~2020B01), the Fundamental Research Funds for the Central Universities (No.~2242019R30001), the EU H2020 RISE TESTBED2 project (No.~872172), and the Open Foundation for Graduate Innovation of NUAA (No.~KFJJ~20180408).

\section*{Acknowledgment}
This work was supported by the National Key R\&D Program of China (No.~2018YF1801101), the National Key Scientific Instrument and Equipment Development Project (No.~61827801), the National Natural Science Foundation of China (No.~61960206006), the Research Fund of National Mobile Communications Research Laboratory, Southeast University (No.~2020B01), the Fundamental Research Funds for the Central Universities (No.~2242019R30001), the EU H2020 RISE TESTBED2 project (No.~872172), and the Open Foundation for Graduate Innovation of NUAA (No.~KFJJ~20180408).

%









\begin{thebibliography}{00}

\bibitem{Zheng17}
K. Zheng, L. Zhao, J. Mei, B. Shao, W. Xiang, and L. Hanzo, ``Survey of large-scale MIMO systems,'' \textit{IEEE Commun. Survey and Tut.},
vol.~17, no.~3, pp.~1738--1760, Apr. 2015.

\bibitem{Liu17_SCIS}
Y. Liu, A. Ghazal, C.-X. Wang, X. Ge, Y. Yang, and Y. Zhang, ``Channel measurements and models for high-speed train wireless communication systems in tunnel scenarios: a survey,'' \textit{Sci. China Inf. Sci.}, vol.~60, no.~10, doi: 10.1007/s11432-016-9014-3, Oct. 2017.


\bibitem{WCX18_Survey}
C.-X. Wang, J. Bian, J. Sun, W. Zhang and M. Zhang, ``A survey of 5G channel measurements and models,'' \textit{IEEE Commun. Surveys and Tut.},
vol.~20, no.~4, pp.~3142--3168, 4th Quart., 2018.
\bibitem{WCX18_TCom}
S. Wu, C.-X Wang, M. Aggoune, M. M. Alwakeel, and X. H. You, ``A general 3-D non-stationary 5G wireless channel
model,'' \textit{IEEE Trans. Commun.}, vol.~66, no.~7, pp.~3065--3078, July 2018.

\bibitem{Ge16_JSAC}
X. Ge, J. Ye, Y. Yang and Q. Li, ``User Mobility Evaluation for 5G Small Cell Networks Based on Individual Mobility Model,'' \textit{IEEE J. Sel. Areas Commun.}, vol.~34, no.~3, pp.~528--541, Mar. 2016.

\bibitem{Makhoul17}
G. Makhoul, F. Mani, R. DErrico, and C. Oestges, ``On the modeling of time correlation functions for mobile-to-mobile fading channels in indoor
environments,'' \textit{IEEE Antennas Wireless Propag. Lett.}, vol.~16, pp.~549--552, Mar. 2017.
\bibitem{Zhu19_WCL}
Q. Zhu, Y. Yang, C.-X. Wang, et al., ``Spatial correlations of a 3D nonstationary MIMO channel model with 3D antenna arrays and 3D arbitrary
trajectories,'' \textit{IEEE Wireless Commun. Lett.}, vol.~8, no.~2, pp.~512--515, Apr. 2019.
\bibitem{Fan16}
W. Fan, I. Carton, J. $\varnothing$. Nielsen, K. Olesen, and G. F. Pedersen, ``Measured wideband
characteristics of indoor channels at centimetric and millimetric bands,''
\textit{EURASIP J. Wireless Commun. Netw.}, vol.~2016, no.~1, pp.~58--70, Feb. 2016.
\bibitem{Guan16}
K. Guan, B. Ai, M. L. Nicolas, et al., ``On the influence of scattering from traffic signs in vehicle-to-x
communications,'' \textit{IEEE Trans. Veh. Technol}, vol.~65, no.~8, pp.~5835--5849, Aug. 2016.
\bibitem{Ge15_TCom}
X. Ge, B. Yang, J. Ye, G. Mao, C.-X. Wang and T. Han, ``Spatial Spectrum and Energy Efficiency of Random Cellular Networks,'' \textit{IEEE Trans. Commun.}, vol.~63, no.~3, pp.~1019--1030, Mar. 2015.
\bibitem{Zhong17_JSAC}
Y. Zhong, T. Quek, X. Ge, ``Heterogeneous Cellular Networks with Spatio-Temporal Traffic: Delay Analysis and Scheduling,'' \textit{IEEE J. Sel. Areas Commun.}, vol.~35, no.~6, pp.~1373--1386, June 2017.


\bibitem{Sangjo18}
S. Yoo, D. Gonzalez, J. Hamalainen, and K. Kim, ``Doppler spectrum analysis of a roadside scatterer model for vehicle-to-vehicle channels: an
indirect method,'' \textit{IEEE Trans. Wireless Commun.}, vol.~17, no.~12, pp.~8007--8021, Dec. 2018.
\bibitem{Liang16_ChinaCom}
X. Liang, X. Zhao, S. Li, Q. Wang, and W. Lu, ``A 3D geometry-based scattering model for vehicle-to-vehicle wideband MIMO relay-based cooperative
channels,'' \textit{IEEE China Commun.}, vol.~13, no.~10, pp.~1--10, Oct. 2016.
\bibitem{Adrian15}
A. Ispas, C. Schneider, G. Ascheid , and R. Thoma, ``Analysis of the local quasi-stationary of measured dual-polarized MIMO
channels,'' \textit{IEEE Trans. Veh. Technol.}, vol~64, no.~8, pp.~3481--3493, Sep. 2014.
\bibitem{Gu18}
C. A. Guti¨¦rrez, J. T. Guti¨¦rrez-Mena, J. M. Luna-Rivera, D. U. Campos-Delgado, R. Vel¨¢zquez, and M. Patzold, ``Geometry-based statistical modeling of non-WSSUS mobile-to-mobile Rayleigh fading channels,''
 \textit{IEEE Trans. Veh. Technol.}, vol.~67, no.~1, pp.~362--377, Aug. 2018.
\bibitem{Liang18_Access}
X. Liang, W. Cao, and X. Zhao, ``Doppler power spectra for 3D vehicle-to-vehicle channels with moving
scatterers,'' \textit{IEEE Access}, vol.~6, pp.~42822--42828, July 2018.

\bibitem{WCX16_TVT}
Y. Fu, C.-X. Wang, Y. Yuan, et al., ``BER performance of spatial modulation systems under 3D V2V MIMO channel models,'' \textit{IEEE Trans. Veh. Technol.}, vol.~65, no.~7, pp.~5725--5730, July 2016.

\bibitem{WCX15_TWC}
Y. Yuan, C.-X. Wang, Y. He, M. M. Alwakeel, and H. Aggoune, ``3D wideband non-stationary geometry-based stochastic models for non-isotropic MIMO vehicle-to-vehicle channels,'' \textit{IEEE Trans. Wireless Commun.}, vol.~14, no.~12, pp.~6883--6895, Dec. 2015.

\bibitem{Ruisi18-TWC}
R. He, B. Ai, G. L. Stber, and Z. Zhong, ``Mobility model based non-stationary mobile-to-mobile channel modeling,''
 \textit{IEEE Trans. Wireless Commun.}, vol.~17, no.~7, pp.~4388--4400, Apr. 2018.
\bibitem{Jiang19_Access}
H. Jiang, Z. Zhang, and G. Gui, ``A novel estimated wideband geometry based vehicle-to-vehicle channel model using an AoD and AoA estimation
algorithm,'' \textit{IEEE Access}, vol.~7, pp.~35124--35131, Feb. 2019.
\bibitem{WCX18_TWC}
J. Bian, J. Sun, C.-X. Wang, et al., ``A WINNER+ based 3D non-stationary wideband MIMO channel model,'' \textit{IEEE Trans. Wireless Commun.},
vol.~17, no.~3, pp.~1755--1767, Mar. 2018.
\bibitem{WCX_Access19}
J. Bian, C.-X. Wang, J. Huang, et al., ``A 3D wideband non-stationary multi-mobility model for vehicle-to-vehicle MIMO
channels,'' \textit{IEEE Access}, vol.~7, no.~1, pp.~32562--32577, Dec. 2019.
\bibitem{Zhu18_Access}
Q. Zhu, Y. Yang, X. Chen, et. al., ``A novel 3D non-stationary vehicle-to-vehicle channel model and its spatial-temporal correlation
properties,'' \textit{IEEE Access}, vol.~6, pp.~43633--43643, July~2018.
\bibitem{Zhu19_CL}
Q. Zhu, W. Li, C.-X. Wang, et al., ``Temporal correlations for a non-stationary vehicle-to-vehicle channel model allowing velocity
variations,'' \textit{IEEE Commun. Lett.}, vol.~23, no.~7, pp.~1280--1284, July 2019.
\bibitem{Hofstetter06_EuCAP}
H. Hofstertter, A. F. Molisch, and N. Czink, ``A twin-cluster MIMO channel model,''
 in \textit{Proc. Antennas and Propagation (EuCAP)}, Nice, France, Nov.~2006, pp.~1--8.
\bibitem{Zhu18_TWC}
Q. Zhu, H. Li, Y. Fu, et al., ``A novel 3D non-stationary wireless MIMO channel simulator and hardware
emulator,'' \textit{IEEE Trans. Commun.}, vol.~66, no.~9, pp.~3865--3878, Sep. 2018.
\bibitem{Pedersen00}
 K. I. Pedersen, P. E. Mogensen, and B. H. Fleury, ``A stochastic model of the temporal and azimuthal dispersion seen at the base
 station in outdoor propagation environments,'' \textit{IEEE Trans. Veh. Technol.}, vol.~49, no.~2, pp.~437--447, Mar. 2000.
\bibitem{Zajic08}
A. G. Zajic and G. L. St¨¹ber, ``Space-time correlated mobile-to-mobile channels: modelling and
simulation,'' \textit{IEEE Trans. Vehic. Tech.}, vol.~57, no.~2, pp.~715--726, Mar.~2008.
\bibitem{Gradshteyn14}
 I. S. Gradshteyn and I. M. Ryzhik. ``Table of integrals, series, and products,'' \textit{Academic press}, 2014.


\end{thebibliography}
\end{document}